# Room Temperature Continuous-wave Excited Biexciton Emission in CsPbBr$_3$ Nanocrystals

*Jie Chen,[1,2,3] Qing Zhang,[3,4]\* Wenna Du,[1] Yang Mi,[1] Qiuyu Shang,[3,4] Jia Shi,[1,2] Pengchong Liu,[1,2] Xinyu Sui,[1,2] Xianxin Wu,[1,2] Rui Wang,[1] Bo Peng,[5] Haizheng Zhong,[6] Guichuan Xing,[7] Xiaohui Qiu,[1,2] Tze Chien Sum,[8]\* and Xinfeng Liu[1,2]\**

[1]Division of Nanophotonics, CAS Key Laboratory of Standardization and Measurement for Nanotechnology, CAS Center for Excellence in Nanoscience, National Center for Nanoscience and Technology, Beijing 100190, China

[2]University of Chinese Academy of Sciences 19 A Yuquan Rd, Shijingshan District, Beijing 100049, China

[3]Department of Materials Science and Engineering, College of Engineering, Peking University, Beijing 100871, China

[4]Research Center for Wide Band Semiconductor, Peking University, China

[5]National Engineering Research Center of Electromagnetic Radiation Control Materials and State Key Laboratory of Electronic Thin Films and Integrated Devices, University of Electronic Science and Technology of China

[6]Beijing Key Laboratory of Nanophotonics and Ultrafine Optoelectronic Systems, School of Materials Science & Engineering, Beijing Institute of Technology, China

[7]Institute of Applied Physics and Materials Engineering, University of Macau, China

[8]School of Physical & Mathematical Sciences, Nanyang Technological University, 21 Nanyang Link, Singapore

\*Email: q_zhang@pku.edu.cn; tzechien@ntu.edu.sg; liuxf@nanoctr.cn

**Biexcitons are a manifestation of many-body excitonic interactions crucial for quantum information and quantum computation in the construction of coherent combinations of quantum states. However, due to their small binding energy and low transition efficiency, most biexcitons in conventional semiconductors exist either at cryogenic temperature or under femtosecond pulse laser excitation. Here we demonstrate room temperature, continuous wave driven biexciton states in $CsPbBr_3$ perovskite nanocrystals through coupling with a plasmonic nanogap. The room temperature $CsPbBr_3$ biexciton excitation fluence (~100 mW/cm$^2$) is reduced by ~$10^{13}$ times in the Ag nanowire-film nanogaps. The giant enhancement of biexciton emission is driven by coherent biexciton-plasmon Fano interference. These results provide new pathways to develop high efficiency non-blinking single photon sources, entangled light sources and lasers based on biexciton states.**

Lead halide perovskites, with their outstanding carrier transport characteristics, high emission quantum yield, tunable bandgaps, and large absorption coefficients[1, 2, 3, 4], have attracted great interests for applications across a range of technologies from solar energy conversion[1, 2, 5] to light-emitting diodes (LEDs)[6]. In perovskites electrons and holes are confined into inorganic $[PbI_6]^{4-}$ octahedral networks, which lead to enormous Coulomb interaction between electrons and holes and therefore strong excitonic effects[7, 8]. In low dimensional perovskite structures *i.e.* nanocrystals (NCs), the electron-hole interaction is further enlarged as a result of increased spatial overlap between electrons and holes. For instance, exciton binding energy of $CsPbBr_3$ perovskite bulk crystals is 40 meV[9], while the value increases to be 120 meV in NCs[10]. Also, CsPbBr$_3$ perovskite NCs perform high quantum yield of ~90%[11]. The large exciton binding energy and high quantum efficiency of CsPbBr$_3$ perovskite NCs holds the key to the development of excitonic and quantum devices with their stable excitons at room temperature.

A biexciton is formed by two free excitons in condensed exciton systems[12], two photon absorption[13], or excitation from single exciton state to biexciton state is of fundamental interests and applicable importance for quantum information and computation for its overwhelming advantage in construction of coherent combination of quantum states. The coherent control of biexciton via its four physically distinguishable quantum states can be applied to basic quantum operation, *i.e.* a two-bit physical quantum and conditional quantum logic gate[14, 15, 16]. If fine structure splitting of exciton state is smaller than natural linewidth, the two indistinguishable radiative decay pathways of a biexciton can render a source of polarization entangled photon pairs. Furthermore, the nonlinear optical nature of biexciton process can generate probabilistic number of entangled-photon pairs per excitation cycle[17], which is similar to standard entangled source parametric down conversion[18] and four wave mixing sources[19]. Moreover, biexciton does not have dark state so that the light sources based on biexciton show no blinking and high saturation intensity[20, 21]. These intrinsic outstanding physical properties make biexciton also interesting for inherent and coherent light sources including lasers, light emitting diode, *etc*[22, 23]. Till now, biexcitons have been realized in semiconductor heterostructures[24], NCs[22] and two dimensional semiconductors[12, 25, 26], *etc*. Due to small biexciton binding energy and large Auger effect, biexcitonic effects have only been realized either under intense pumping by short pulsed lasers (~MW/cm$^2$) or cryogenic condition, which severely limits their practical applications. Resonant excitation can increase probability of biexction transition. For example, in quantum wells biexciton emission was observed under a continuous wave (CW) excitation beam with power of ~70 mW/cm$^2$, but only below liquid helium temperature due to small biexciton binding energy[27, 28]. So far, continuous wave (CW) excitation, room temperature operation of biexciton with fundamental applicable significance is extremely challenging.

Plasmonic nanogap structures have been extensively explored in quantum electromagnetic dynamics (QED) due to its extraordinary capability of tailoring strong and weak light-matter interaction in deep-subwavelength regime[29]. In the last decade, plasmonic nanogap is widely used to enhance a variety of linear and nonlinear optical processes including emission, Raman, and high harmonic generation, *etc*[30, 31, 32]. The interactions between plasmon and quantum dots are widely discussed[33, 34, 35], where the surface plasmon is regarded as continuous energy states while quantum dots excitations are discrete energy levels. The coupling between continuous energy states and discrete energy states leads to Fano effect[36]. In this work, we demonstrate room temperature, continuous wave excited biexciton states in CsPbBr$_3$ perovskite nanocrystals through coupling with the plasmonic nanogap. The room temperature CsPbBr$_3$ biexction excitation fluence is reduced by ~$10^{13}$ times with Ag nanowire-film nanogap. The giant enhancement of biexciton emission is dominantly induced by coherent biexciton-plasmon Fano resonance. These results provide new pathways to develop high efficiency non-blinking single photon sources, entangles light sources and lasers based on biexciton states.

**Figure 1a** shows the energy level diagrams involving excitation and decay of a single exciton (left panel) and a biexciton (right panel). Single exciton bright state can decay to the ground state by emitting one photon while dark state decays to ground state by nonradiative process. Two states are coupled by spin-flip process. Biexciton can decay to either of two single exciton states (bright state) accompanied by emission of one photon. The biexciton emission energy $\hbar\omega_{xx}$ is determined by energy gap between biexciton energy $E_{xx}$ and single exciton emission $E_x$ via $\hbar\omega_{xx}=E_{xx}-E_x$. Therefore, when two excitons bind together to form one biexciton, the energy of whole system decreases by $\Delta_{xx} = \hbar\omega_x - \hbar\omega_{xx}$, which is so-called biexciton binding energy. **Figure 1b** shows the plasmon enhance biexciton state three processes. (I) Nanocrystal absorb one photon to exciton state; (II) Biexciton state and nanostructure coupled together; and (III) Nanostructure transfer the energy to exciton and plasmon relax to ground state while exciton reach biexciton state. Our designed structure is shown in **Figure 1c**, which is a sandwiched structure of silver nanowire/NCs/silver film. A thin layer of $SiO_2$, perovskite NCs are coated onto silver film step by step. The $CsPbBr_3$ nanocrystals (*Supplementary Information*, **Figure S1**) show good crystallinity with a cubic structure (*Supplementary Information*, **Figure S2**). $SiO_2$ is used to control gap width $d_G$ between Ag nanowire and film and also prevent perovskite NCs from photoluminescence quenching of perovskite NC emitters nearby metal surface. To reduce scattering and radiation losses in plasmonic nanogap due to surface discontinuity, growth condition of $SiO_2$/Ag film is optimized with surface roughness ~1.1 nm (*Supplementary Information*, **Figure S3**). **Figure 1d** shows schematics of biexcitons in the plasmonic nanogap. Perovskite NCs deposited onto $SiO_2$ layer by solution processed spin-coating methods. The thickness of perovskite NCs layer is 13 nm measured by ellipsometry (*Supplementary Information*, **Figure S4**). Cross-section scanning electron microcopy (SEM) image (**Figure 1e**) shows that the perovskite NCs layer is uniform in scale of millimeters.

To probe biexciton effect in perovskite NCs, photoluminescence (PL) of perovskite NCs on and off plasmonic nanogaps is performed with a continuous wave laser (**Figure 2a**, wavelength: 405 nm; spot diameter: 2 μm). PL spectroscopy far away from plasmonic nanogap ($P_1$, blue line) shows single symmetric peak located at 500 nm, indicated as free exciton as excitation power $P$ is 1.6 μW as shown in **Figure 2a**, which is assigned to single exciton recombination of $CsPbBr_3$[37]. As a contrast, on the nanogap ($P_2$, red line), a new peak arises at 520 nm with intensity much higher than single exciton emission (linear coordinate PL spectrum is shown in *Supplementary Information*, **Figure S5**). To confirm plasmon mode between silver nanowire and film, emission polarization behavior is measured. As shown in **Figure 2b**, biexciton emission is strongest when excitation polarization direction is parallel to long axis of nanowires (0°, 180°, 360°) and weakest when excitation polarization direction is perpendicular to long axis (90°, 270°). However, the emission intensity is almost the same away from the Ag nanowire (*Supplementary Information*, **Figure S6**). The polarization resolved emission properties of 520 nm peak suggests the modulation of Ag nanowire plasmonic modes which is transverse magnetic mode. The energy difference between single exciton (500 nm, 2.480 eV) and new emission peak (520 nm, 2.385 eV) is ~ 95 meV, which is almost the same as biexciton binding energy of $CsPbBr_3$ NCs reported previously (~100 meV)[23]. Temperature dependent PL spectroscopy was also performed to verify the binding energy of $CsPbBr_3$ NCs. As shown in **Figure 2c**, PL increases as temperature decreases from 295 to 80 K (*Supplementary Information*, **Figure S7**), which could be well fitted by a Boltzmann distribution function with a thermal active energy $E_b$ of ~100 meV (**Figure 2c**). The thermal active energy of 520 nm peak $E_b$ is almost equal to energy difference between it and free exciton emission, strongly indicating that the 520 nm peak is ascribed to biexciton emission according to **Figure 2a**. The biexciton binding energy is much larger than room temperature thermal fluctuation energy, supporting that the biexciton is stable and observable at room temperature.

Under full thermal equilibrium conditions of the biexciton recombination process, emission intensity of biexciton $I_{bx}$ is proportional to the square of single exciton emission intensity $I_{ex}$[38,39]. Consequently, biexciton lifetime is about half of single exciton lifetime. Next, excitation power dependent PL and time resolved PL spectroscopy is conducted to confirm the occurrence of biexciton. **Figure 3a and b** shows power dependent PL spectroscopy on and off silver nanowire on Ag film, respectively. **Figure 3c** shows the integrated emission intensity of exciton ($I_{ex}$) and biexction ($I_{bx}$) as a function of the fluence intensity *P*. As pumping power increases from 20 μw to 130 μw, PL spectroscopy away from plasmonic nanogap ($P_1$, **Figure 2**) show one emission peak at 500 nm with intensity *I* linear with pumping power *P* (**Figure 3c**, purple), which is ascribed to single exciton recombination. However, on plasmonic nanogap 520 nm peak due to biexciton emission emerges with intensity $I_{bx}$ grows much faster that of single exciton (**Figure 3b**). Although the biexciton emission peak onset emerges when $P > 100$ mW/cm$^2$ (*Supplementary Information*, **Figure S8**), we hereby performed the emission spectroscopy with $P > 500$ W/cm$^2$ (20 μW) to lower the fitting error in the separating the power dependence of the biexction emission from that of the single exciton emission. **Figure 3c** shows the integrated emission intensity of exciton ($I_{ex}$) and biexction ($I_{bx}$) as a function of the fluence intensity. The power dependence of biexciton emission can be described adequately by superlinear function with a power-law of $k$ =1.83. The power law of biexciton (1.83) is about 1.91 times of that for single exciton (0.96), strongly suggesting the occurrence of biexciton emission.

Further, time-resolved PL spectroscopy is conducted to probe the exciton and biexciton dynamics. Quantitatively under thermal equilibrium condition, the time evolution of the exciton and biexciton can be modeled with the transition functions[40] $n_{bx} \approx n_{bx}^0 e^{-2t/\tau_{ex}}$, when exchange interaction between excitons with different spin is ignored. In the other words, biexction decay rate indeed about twice of single exciton, as suggested in bulk semiconductor [41], quantum dots[40], quantum wells[42] and 2D semiconductors[43]. **Figure 4a-b** shows TRPL spectra of CsPbBr$_3$ on (**a**) and off (**b**) plasmonic nanogap. To avoid any biexciton effects away from plasmonic nanogap under high pumping conditions, the excitation fluence of femto-second pulsed laser is ~0.1 nJ/cm$^2$ and thereby PL spectra exhibit only single exciton emission. The single exciton recombination curve can be well fitted by a single exponential decay function with time constant of $\tau_{ex}$ = 1170 ± 10 ps. The TRPL spectroscopy of biexciton emission can also be well-fitted by a single exponential decay curve, suggesting the emission on plasmonic nanogap is dominated by biexciton recombination. The lifetime of biexciton $\tau_{be}$ = 550 ± 5 ps, is half of single exciton lifetime (1170 ± 10 ps), which is consistent with transition properties of biexcition states as discussed above. Therefore, we can shortly conclude that the 520 nm peak arising on plasmonic nanogap is due to biexction recombination.

Due to large binding energy of CsPbBr$_3$ NCs, room temperature biexciton emission off nanogap can be excited only by femto-second laser and the pumping fluence is in the magnitude of ~$10^{12}$ W/cm$^2$ (see *Supplementary Information*, **Figure S9**). In the plasmonic nanogap, biexciton emission is observed with the pumping fluence exceeds 100 mW/cm$^2$ (see *Supplementary Information*, **Figure S8**). Therefore, the biexciton carrier density threshold in the plasmonic nanogap is reduced by ~$10^{13}$ times (*Supplementary Information*, **Figure S9**). However, the local field enhancement factor, which is much larger than local field enhancement factor (<100, *Supplementary Information*, **Figure S10**) in the plasmonic nanogap. It means that the enhanced biexciton emission is not due to local field enhancement. Biexciton forms mainly through the following three methods: 1) two photon absorption; 2) exciton-exciton bounding in condensed exciton environment; 3) resonant excitation from single exciton. The first two pathways are strongly required intense pumping, which is not feasible under continuous wave excitation conditions. Further, scattering spectroscopy is conducted to explore exciton and plasmon resonant energy transfer. Scattering spectra of bare Ag nanowire shows broad scattering peak when the detection spectra range is 450 to 750 nm (*Supplementary Information*, **Figure S11**), which depends on the nanowire diameter, as reported in previous literatures[44]. Moreover, an asymmetric peak appears around perovskite biexciton resonant energy, resulting from Fano interference between biexciton resonance excitation and background continuum states excitation of surface plasmon modes as shown in **Figure 5a**. The fitted Fano factor is 0.25 and 0.2 respectively, when polarization is parallel and vertical to long axis of nanowire and simulated with FDTD simulation in **Figure 5b**. It is noticed that Fano resonance lineshape is only observed in scattering spectra of Ag nanowire-NCs-Ag film system with strong biexciton emission, which suggests that the exciton-plasmon energy transfer plays a vital role in the formation of biexciton. Plasmon resonant energy transfer can only happen when the plasmon peak overlapped with biexciton states (*Supplementary Information*, **Figure S11**). Without energy transfer, NCs in the gap only show exciton emission (*Supplementary Information* **Figure S12**). A resonant excitation process is proposed to explain the huge biexciton effect driven by surface plasmon. As shown in **Figure 1b**, with optical pumping electrons are to single exciton states of perovskite NCs and surface plasmon states above Fermi level of silver, which in resonant with biexciton states. The interference of the two excitation processes result in the Fano resonance lineshape in the scattering spectra. The electrons in surface plasmon band and biexciton states interfere with each other. Since single exciton-biexciton transition energy is nearly as the same as SP, resonant energy transfer from SP to single exciton occurs, leading to the excitation of exciton and annihilation of SP. The surface plasmon driven excitation mechanism is quite similar as that as-demonstrated resonant pump-probe biexciton in single carbon nanotube[45]. The resonance effect extensively promoted the excitation cross-section of biexciton from single exciton states.

**Conclusions**

In conclusion, we have observed clear evidence for the presence of high efficiency biexciton in perovskite NCs in metallic nanostructure under continuous wave pump at room temperature. The observation of the biexciton is a consequence of the very clear emission spectra besides exciton with a binding energy ~95 meV and verified by activation energy. The power law of fluorescence and lifetime give further evidence to biexciton emission. We further verified the Fano effects by measuring scattering spectrum. By fitting the spectrum with Fano parameters and simulated with FDTD, we confirm that this demonstration of a four-particle complex in plasmonic nanostructure will give rise new interesting effects, such like quantum logic gates, Bose-Einstein-condensation of exciton, source of polarization-entangled photons and single-photon source.

**Methods**

***Sample preparation:*** The SiO$_2$/Silver substrate was made though magnetron sputtering with silver thickness 50 nm and SiO$_2$ thickness 5 nm on SiO$_2$/Si substrate. CsPbBr$_3$ NCs were spin coating on the SiO$_2$/Silver substrate with a speed 3000 r/min. The 4-Methyl-1-acetoxycalix[6]arene (4M1AC6, 0.5% in chlorobenzene) was spin coating on CsPbBr$_3$ NCs with a speed 5000 r/min. Waiting for about half an hour to allow 4M1AC6 to dry and then spin coat silver nanowires in isopropyl alcohol with a speed 2000 r/min.

***SEM and TEM measurement***: The samples were force apart after liquid nitrogen treatment. The SEM equipment was Merlin-61-53 with a work distant 4.3 mm and voltage of 5 KV. The perovskite NCs for transmission electron microscope (TEM) measurement were dropped onto the TEM grids. High resolution transmission electron microscopy (HRTEM) was done with a FEI Tecnai F20 operated with an acceleration voltage of 200 kV.

***Time-resolved Photoluminescence Spectroscopy***: For time-resolved photoluminescence measurements, the excitation pulses (wavelength 400 nm) are doubled frequencies of a Coherent Mira 900 (120 fs, 800 nm, 76 MHz) and filtered by a 655 short pass to generate 400 nm and the backscattered signal was collected using a Time Correlated Single Photon Counting (TCSPC, SPC-150) which has an ultimate temporal resolution of ~40 ps. A 442 nm long pass filter is placed before the optical fiber to filter out the residual 400 nm.

***PL measurement***: The output from a 405 nm continuous wave laser circularly polarized by the quarter-wave plate, and focused on a sample by microscope objective lens (100×, NA = 0.95, with spot size ~2 μm). The PL signal was then back collected by the same lens, and filtered by a long pass filter before entering a spectrometer (PI Acton 2500i with a liquid nitrogen cooled charge coupled device − CCD camera). The low temperature PL was measured when the sample was in liquid nitrogen refrigeration cryogenic instrument

***FDTD simulation***: The FDTD simulation of silver nanowire and quantum dots is simulated with Lorentz model[44]. We treat CsPbBr$_3$ nanocrystal layer modeling the dielectric function as a single Lorentzian function. The dielectric constant for bulk $CsPbBr_3$ is $3.8^{46}$. Biexciton energy and plasmon resonant energy transfer are considered while other high order transitions are ignored. Biexciton linewidth are used as transition linewidth. The metal dielectric function is treated as a Lorentz-Drude model.

**Associated content**

*Supplementary Information

The Supplementary Information is available.

Figure S1-S12 as described in the text.

**Author information**

Corresponding Authors

*E-mail: q_zhang@pku.edu.cn; tzechien@ntu.edu.sg; liuxf@nanoctr.cn

**Author contributions**

X.L, Q.Z conceived the idea for the manuscript and designed the experiments. J.C, W.D, Y.M, Q.S, J.S and R.W conducted the spectroscopic characterization. B.P and H.Z and P.L prepared the perovskite QDs. J.C, W.D fabricated devices and performed TEM, SEM and AFM. J.C. performed the simulations. X.L, Q.Z, T.C. S, X.Q and G.X contributed to the data analysis. All the authors discuss the results and the manuscript. Q.Z, T.S and X.L led the project.

**Acknowledgements**

The authors thank the support from the Ministry of Science and Technology (2017YFA0205700, 2017YFA0304600, 2016YFA0200700 and 2017YFA0205004), National Natural Science Foundation of China (No. 21673054). Q.Z. acknowledges the support of start-up funding from Peking University, one-thousand talent programs from Chinese government, open research fund program of the state key laboratory of low-dimensional quantum physics. T.C.S. acknowledges the financial support from the Ministry of Education

(MOE) AcRF Tier 1 grants RG101/15 and RG173/16; and MOE Tier 2 grants MOE2015-T2-2-015 and MOE2016-T2-1-034; and from the Singapore National Research Foundation through the Competitive Research Program NRF-CRP14-2014.

**Competing financial interests**

The authors declare no competing financial interests.

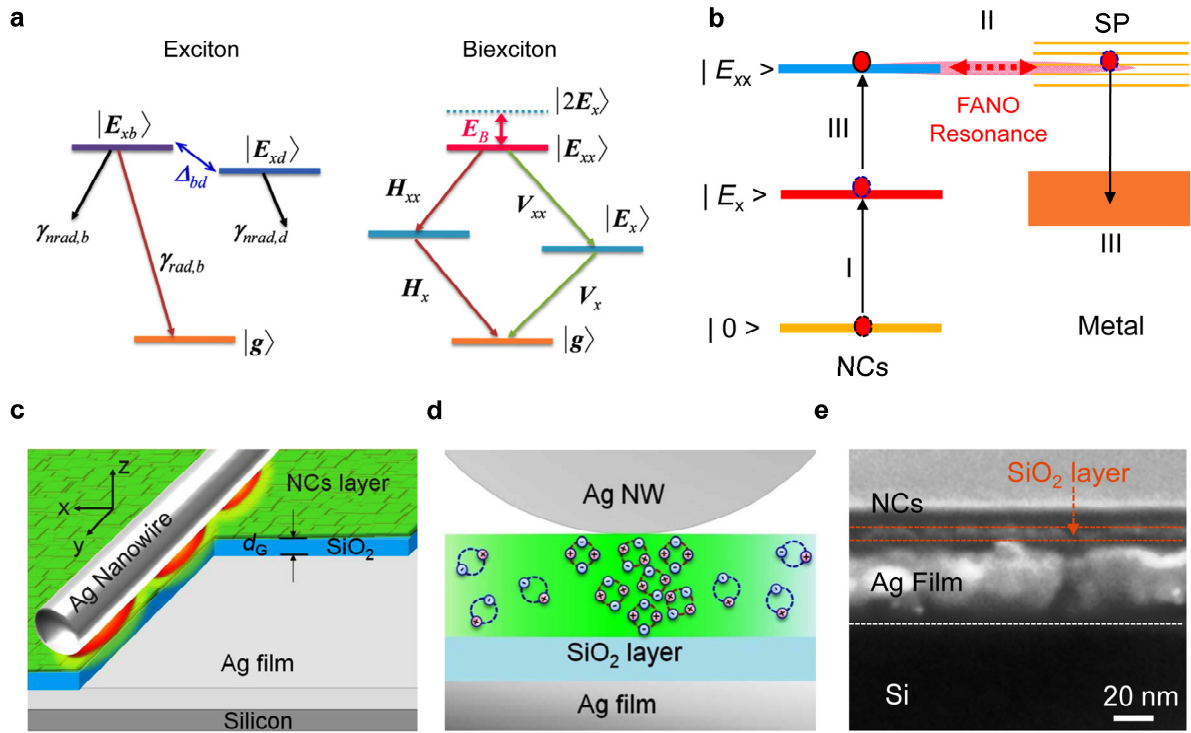

**Figure 1| Schematic view of biexciton and formation process.** (**a**) Schematic diagram of exciton and biexciton transition. Exciton has a bright state which can arrive in ground state by radiation or non-radiation transition. The corresponding dark state coupled though spin-flip process can arrive in ground state by non-radiation transition. The biexciton levels are formed by Four-level system, biexciton level, two bright exciton levels, and ground state. The two bright exciton levels are separated by fine structure splitting. The biexciton decays through a cascaded process of either emitting two horizontally or two vertically polarized photons. (**b**) Schematic diagram of biexciton formation process with FANO effect. (I) Nanocrystal absorb one photon to exciton state. (II) Biexciton state and nanostructure coupled together. (III) Nanostructure transfer the energy to exciton and plasmon relax to ground state while exciton reaches biexciton state. (**c**) Schematic view of structure. The cavity is composed of a silver nanowire and a silver substrate separated by a gap of 5 nm $SiO_2$ and perovskite NCs. (**d**) Schematic view of the exciton and biexciton in the cavity. (**e**) Cross-section SEM of the sample.

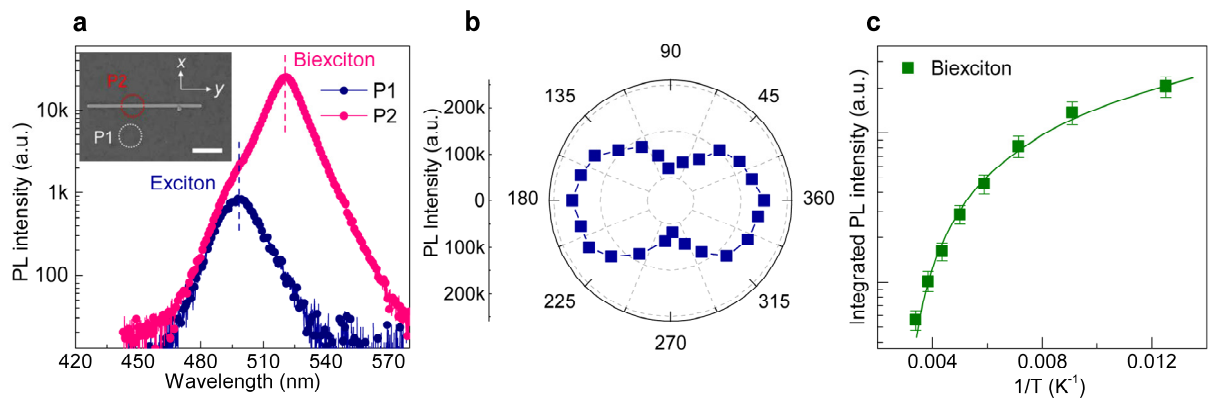

**Figure 2| Spectral characteristics of cavity-coupled perovskite NCs fluorescence**. (**a**) Fluorescence spectra of Perovskite NCs off-cavity ($P_1$) shows single symmetric exciton peak at 500 nm while on-cavity ($P_2$) a biexciton peak at 520 nm under continuous wave radiation emerge. The binding energy of biexciton is about 95 meV. Inset is SEM image of a cavity (scale bar 1 μm), illustrating the location of the incident laser spot for on-cavity ($P_2$) and off-cavity ($P_1$) measurements. (**b**) Polarization of biexciton shows the biexciton is along the silver nanowire. (**c**) Thermal stability of the biexciton using temperature dependent PL measurements, with fitted thermal activation energies of around 100 meV.

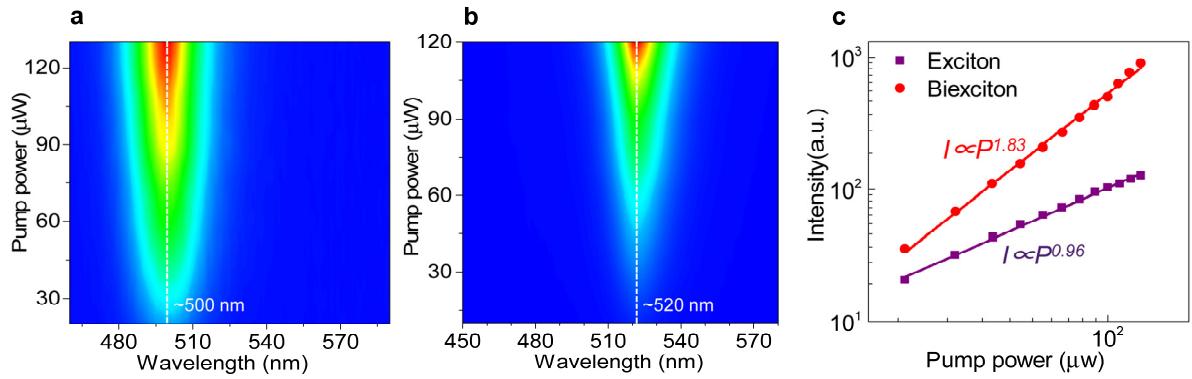

**Figure 3| Photoluminescence spectra and integral intensities of perovskite NCs for different pump fluences**. (**a**) PL spectroscopy off the nanogap only show exciton recombination peaks as pumping power increased from 20 μw to 130 μw. The emission intensity is linear with pumping power (**c**, purple). (**b**) The biexciton emission peak of 520 nm at high excitation fluence. With increasing pumping power, the biexciton, however, grows much faster that of single exciton. Same color map is used to make a clear comparison. (**c**) Log-log scale is adopted to plot the integrated emission intensity strength, $I_p$, as a function of the exciton emission strength $P$. The purple line is a power-law fit $I_p=P^k$, with $k_1$=0.96 while the red line $k_2$=1.83, the ratio is derived $l=k_2/k_1$, with $l$=1.91. This also gives a strong evidence that (**a**) is exciton luminescence while (**b**) is biexciton fluorescence.

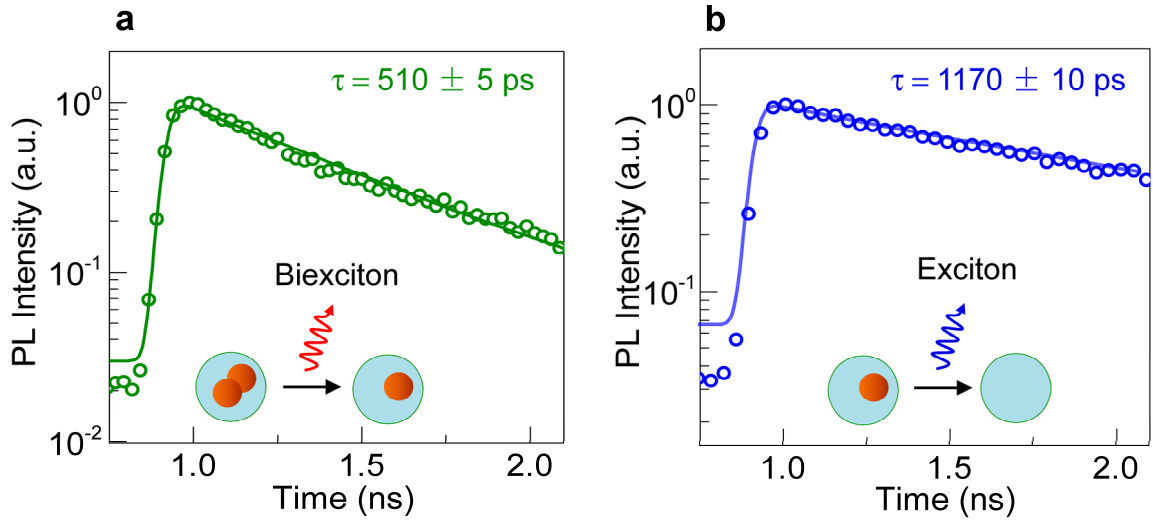

**Figure 4| Dynamics of excited states**. (**a**) Time-resolved PL of biexciton which can be seen as single decay process shows lifetime is $\tau_{be}$ = 510 ± 5 ps. (**b**) Single index decay of exciton and fitted with a lifetime of $\tau_{ex}$ = 1170 ± 10 ps represents the electron hole recombination dominate the photoluminescence. The lifetime of the exciton is approximately twice of the biexciton. Here, the extracted ratio $B$ = 2.29 is consistent with the theoretical result.

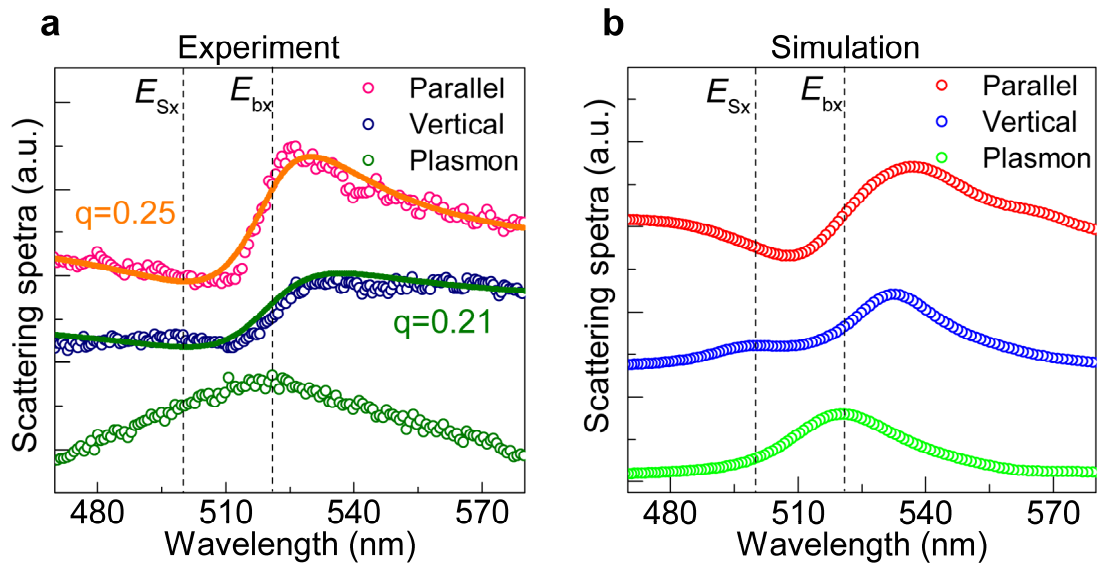

**Figure 5| Fano effect of silver nanowire and perovskite NCs.** (**a**) Plasmon resonant spectra of silver nanowire is a single symmetrical peak shown as green dot line. The silver nanowire with perovskite NCs shows a symmetrical peak. Red and blue dot line shows the scattering spectra with parallel and vertical polarization respectively. (**b**) FDTD simulation of (**a**).